# Approach combining the Rietveld method and pairs distribution function analysis to study crystalline materials under high-pressure and/or temperature

J. C. de Lima[1,a)], Z. V. Borges[1], C. M. Poffo[2], S. M. Souza[3], D. M. Trichês[3], and R.S. de Biasi[4]

[1]*Departamento de Engenharia Mecânica, Universidade Federal de Santa Catarina, Programa de Pós-Graduação em Ciência e Engenharia de Materiais, Campus Universitário Trindade, C.P. 476, 88040-900 Florianópolis, Santa Catarina, Brazil*

[2]*Universidade Federal de Santa Catarina, Campus de Araranguá, 88900-000 Araranguá, Santa Catarina, Brazil.*

[3]*Departamento de Física, Instituto de Ciências Exatas, Universidade Federal do Amazonas, 69077-000 Manaus, Amazonas, Brazil*

[4]*Seção de Engenharia Mecânica e de Materiais, Instituto Militar de Engenharia, 22290-270 Rio de Janeiro, RJ, Brazil*

ABSTRACT

An approach combining the Rietveld method and pairs distribution function analysis was developed and can help to understand the effect of temperature and/or high-pressure on crystalline materials. It was applied to orthorhombic $Ta_2O_5$ compound, and the obtained results were compared with the experimental pairs distribution function reported in the literature, and an excellent agreement was reached. The approach permitted to simulate average pairs distribution functions $G_{Ta-Ta}(R)$, $G_{Ta-O}(R)$, and $G_{O-O}(R)$ and the average radial distribution functions $RDF_{Ta-Ta}(R)$, $RDF_{Ta-O}(R)$, and $RDF_{O-O}(R)$. The coordination numbers for the first neighbors were obtained, and used in the expression to calculate the Cowley-Warren chemical short-range order parameter $\alpha^{CW}$. The calculated value suggests a preference for forming homopolar pairs in the first coordination shell. Previous study on the effect of high-pressure on orthorhombic $Ta_2O_5$ showed a pressure-induced amorphization process. This amorphization can be associated with the preference for forming homopolar pairs in the first coordination shell.

a) Author to whom correspondence should be addressed. Electronic mail: joao.cardoso.lima@ufsc.br or jcardoso.delima@gmail.com

## I. INTRODUCTION

With the development of the synchrotron radiation sources and the diamond anvil cell (DAC), researches in the materials science field using high-pressure has been growing rapidly. This is corroborated by the increasing number of articles reported in the literature in the last two decades. New crystallographic phases in older materials and new materials with specific properties for technological applications have been discovered.

Despite the extraordinary progress in the development of DACs, allowing that several dozen GPa be achieved, they have yet a limited window of few tens (two or three) of degrees, imposing a restriction in the angular range ($2\theta$) in which the X-ray diffraction (XRD) patterns are measured. This limitation makes difficult to visualize the effect of high-pressure on the structure being studied, as well as it increases the difficulties in the determination of new high-pressure structure phases emerging due to the few number of diffraction peaks recorded.

Trying to overcome the difficulties mentioned above, at least partially, we implemented an approach combining the Rietveld method (RM) [1] and pairs distribution function (PDF) analysis [2,3]. RM has been implemented in several computational packages, for example, the GSAS package [4], which has widely been employed to refine and/or to determine structures of the crystallographic phases of single crystal and polycrystalline solids. The theoretical description of PDF analysis as well as its applications to the liquid and amorphous materials is well documented in the literature [2,3,5], not being necessary to be repeated here. The approach to be presented in this study requires only the knowledge of the structure of crystalline material. It has already been used by us to investigate the effects of moderate pressures (MPa) and high-pressures (GPa) on nanocrystalline materials [6,7]. In order to corroborate the validity of this approach and also make it more explicit for readers, we choose to apply it to the $Ta_2O_5$ compound previously studied using XRD, RM, and PDF [8,9]. Thus, the results obtained using this approach can be compared with those reported in Refs. [8,9]. This paper reports the results reached, and it is divided in three parts: Introduction, Brief summary of approach, Results and discussion, and Conclusions.

## II. BRIEF SUMMARY OF APPROACH

From a PDF analysis point of view, the structure of liquid, amorphous and crystalline materials with $N$ components is described by $N(N+1)/2$ partial pair correlation functions $G_{ij}(R)$, which are the Fourier transformation of the same number of the partial structure factors $S_{ij}(K)$. The total structure factor $S(K)$ derived from the *X*-ray and/or neutron scattering data is a weighted sum of $S_{ij}(K)$. Here, $K = 4\pi sin\theta/\lambda$ is the transferred momentum, $\lambda$ is the wavelength, and $\theta$ is the diffraction angle. For liquid and amorphous materials, $S_{ij}(K)$ factors are obtained from the deconvolution of $S(K)$ factors. The procedure to obtain both $S(K)$ and $S_{ij}(K)$ factors is well documented in the literature [2,3,5], not being necessary to repeat here.

With respect to application of PDF analysis to study single crystal or polycrystalline materials, $S(K)$ is also obtained from the XRD or neutron diffraction measurements. Although $S(K)$ is a weighted sum of $S_{ij}(K)$, usually $S_{ij}(K)$ are not obtained maybe due to the chemical long-range order in these materials, and also that this can be a hard task. $G(r)$ is obtained by a Fourier transformation of $S(K)$. The procedure to obtain both $S(K)$ and $G(r)$ is also well documented in the literature [2,3,5], not being necessary to repeat here.

When crystalline materials are submitted to high-temperature and/or high-pressure, the knowledge of $S_{ij}(K)$ can be very useful to analyze the effects of these two parameters on the structure and, consequently, on their physical properties. Also, knowledge of $S_{ij}(K)$ can help to understand how the possible phase transformations start. Thus, below, we describe the steps necessary to get the $S_{ij}(K)$ factors for crystalline materials.

1) From the measured XRD and/or neutron patterns to determine the structure (space group, lattice parameters, and atomic coordinates of atoms at the Wickoff positions) using the Rietveld method;

2) To use these structural data as "input data" in a crystallographic software, for example, *Crystal Office 98*® from the Atomic Softek - Canada, and using the option (tool "*output + coordinates + shell structure*") the shell structure around a specific atom " *i* " placed at the origin is obtained. From this shell structure, the coordination numbers $N_{ij}$ and the interatomic distances $R_{ij}$ of all other atoms with respect to specific atom " *i* " are obtained. If the structure has the same specific atom occupying different Wickoff positions, it is necessary to obtain the shell structure corresponding to each Wickoff position. This is done by placing each atom at origin. An example will be shown later.

3) To simulate a $S_{ij}(K)$ factor corresponding to a each shell structure, the coordination numbers $N_{ij}$ and interatomic distances $R_{ij}$ are used in an expression given by Mangin [10] slightly modified by us and described below is used.

$$S_{ij}(K) = 1 + \sum_{R_{ij}} \left( N_{ij} \sum_{K=0}^{K\max} \frac{\sin(KR_{ij})}{KR_{ij}} \right) \quad (1)$$

For each pair of values $N_{ij}$ and $R_{ij}$, the sum must be performed over all the values of $K$ between $K = 0$ and $Kmax$. If there are more than one $S_{ij}(K)$ for the same specific atom " $i$ ", an average $<S_{ij}(K)>$ factor should be obtained. According to Faber and Ziman formalism [5], the $S_{ij}(K)$ factors oscillates around 1.

4) The total structure factor $S(K)$ is obtained as follows:

$$S(K) = \sum_{i,j=1}^{2} W_{ij}(K) S_{ij}(K), \quad (2)$$

$$W_{ij}(K) = \frac{c_i c_j f_i(K) f_j(K)}{<f(K)>^2}, \quad (3)$$

$$<f^2(K)> = \sum_{i=1}^{2} c_i f_i^2(K), \quad (4)$$

and

$$<f(K)>^2 = \left[ \sum_{i=1}^{2} c_i f_i(K) \right]^2, \quad (5)$$

where $c_i$ is the concentration of atoms of type $i$ and $f_i(K) = f_i^0(K) + f'(E) + if''(E)$ is the atomic scattering factor, in electrons unit. For the neutral atom, values of $f^0(K)$ can be calculated using the analytic function given by Cromer and Mann [11]. Away from the $K$-edges of the elements, the real and imaginary parts $f'$ and $f''$ given in table compiled by Sasaki [12] can be used. However, near the $K$-edges of the elements, values of $f'$ and $f''$ must be calculated, for example, using the procedure given in Ref. [13,14].

5) The reduced total distribution functions $\gamma(R)$ and reduced partial distribution functions $\gamma_{ij}(R)$ are related to $S(K)$ and $S_{ij}(K)$ by Fourier transformation and they are written as

$$\gamma(R) = (2/\pi)\int_0^\infty K[S(K)-1]\sin(KR)dK, \qquad (6)$$

$$\gamma_{ij}(R) = (2/\pi)\int_0^\infty K[S_{ij}(K)-1]\sin(KR)dK. \qquad (7)$$

The total pairs distribution function $G(R)$ and partial pairs distribution function $G_{ij}(R)$ are related to the $\gamma(R)$ and $\gamma_{ij}(R)$ through the expressions,

$$\gamma(R) = 4\pi\rho_0 R[G(R)-1], \qquad (8)$$

$$\gamma_{ij}(R) = 4\pi\rho_0 R[G_{ij}(R)-1]. \qquad (9)$$

By considering the maxima of the coordination shells (peaks) in the $\gamma_{ij}(R)$ or $G_{ij}(R)$ function, the average interatomic distances for the first, second, etc, neighbors are obtained.

6) The total radial distribution function $RDF(R)$ and partial radial distribution function $RDF_{ij}(R)$ are written as

$$RDF(R) = 4\pi\rho_0 R^2 G(R), \qquad (10)$$

$$RDF_{ij}(R) = 4\pi\rho_0 c_j R^2 G_{ij}(R). \qquad (11)$$

where $\rho_0$ is the atomic density of crystalline material, in atoms/Å$^3$. The coordination numbers for the first, second, etc, neighbors are obtained by integrating the corresponding coordination shell (peak) in the $RDF_{ij}(R)$ function.

## III. APPLICATION TO ORTHORHOMBIC TANTALUM OXIDE $Ta_2O_5$

As mentioned previously in the section **I**, we will use the structural data listed in Ref. [1] for the orthorhombic Tantalum Oxide ($Ta_2O_5$) compound [8,9]. According to Joint Committee on Powder Diffraction Standards (JCPDS) [15], $Ta_2O_5$ can also crystallize in the monoclinic, anorthic, tetragonal, and hexagonal structures. For the orthorhombic phase, the JCPDS, code PDF 79-1375, gives the values $a$ = 43.997 Å, $b$ = 3.894 Å and $c$ = 6.209 Å, $\alpha = \beta = \gamma = 90°$ for the lattice parameters. The Inorganic Crystal Structure Database (ICSD) [16], code 66366, gives the Wickoff positions for thirteen Tantalum atoms distributed in the sites *1a*, *2e*, and *1c*, and the positions for thirty four Oxigens atoms distributed in the sites *1b*, *2f*, *1d*, *1a*, *2e*, and *1c*. These structural data were used as input data for the *Crystal Office 98®* software, and thirteen shell structures up to $R$ = 25 Å around the Ta atoms were obtained. These shell structures give the coordination numbers $N_{Ta\text{-}Ta}$, $N_{Ta\text{-}O}$ and the interatomic distances $R_{Ta\text{-}Ta}$ and $R_{Ta\text{-}O}$. Also, thirty four shell structures up to $R$ = 25 Å around the O atoms were obtained. These shell structures give the coordination numbers $N_{O\text{-}O}$ and the interatomic distances $R_{O\text{-}O}$. The values $N_{ij}$ and $R_{ij}$ were used in expression (1) to obtain thirteen $S_{Ta\text{-}Ta}(K)$, thirteen $S_{Ta\text{-}O}(K)$, and thirty four $S_{O\text{-}O}(K)$. Average $<S_{Ta\text{-}Ta}(K)>$, $<S_{Ta\text{-}O}(K)>$, and $<S_{O\text{-}O}(K)>$ factors were obtained. In order to reproduce the angular range reached using a DAC, a conventional source (Cu and Mo target), and a synchrotron source, the values of *Kmax* 4, 8, 16, 30 and 60 Å$^{-1}$, with a step $\Delta K$ = 0.025 Å$^{-1}$, were considered to obtain the $S_{ij}(K)$ factors. The ponderation weights were generated using the expression (3), and the $S(K)$ factors for each value *Kmax* was obtained using the expression (2).

The $G_{ij}(R)$, $G(R)$, $RDF_{ij}(R)$, and $RDF(R)$ functions were obtained by Fourier transformation of $S(K)$ and $S_{ij}(K)$ factors. The resolution of these functions is given by the expression $\Delta R=3.8/Kmax$ [2], and the values 0.95 Å for *Kmax* = 4 Å$^{-1}$, 0.475 Å for *Kmax* =

8 Å$^{-1}$, 0.253 Å for $Kmax$ = 15 Å$^{-1}$, 0.127 Å for $Kmax$ = 30 Å$^{-1}$, and 0.063 Å for $Kmax$ = 60 Å$^{-1}$ were calculated. With exception of the Crystal Office 98® software, all the computational codes used were developed by the author.

**IV. RESULTS AND DISCUSSION**

Just for the purpose of illustration, Fig. 1 shows two partial output files of shell structures obtained with the Crystal Office 98® software: one around a Ta atom at origin and another around an O atom at origin. Although all the shell structures were calculated up to $R$ = 25 Å, the output files shown were cut at ≈ 3.46 Å and ≈ 2.79 Å.

Figure 2 shows the $G(R)$ functions as a function of pressure for $Ta_2O_5$ reported in Refs. [8,9]. They were obtained by Fourier transformation of $S(K)$ fators derived from the XRD measurements. It is interesting to note that up to 8.5 GPa the PDF functions are similar, whereas for larger pressures they are well different. Those researcher attributed the difference among them to the presence of a pressure-induced amorphization process. The $G(R)$ function for pressure of 1 atm shows all coordination shells well defined, as it is expected for a solid with chemical long-range order. For this reason, it will be used to compare all the simulated $G(R)$ functions in this study.

Figure 3 shows the simulated average $G(R)$ functions for the values $Kmax$ = 4, 8, 15, 30, and 60 Å$^{-1}$. A comparison between the experimental $G(R)$ function at 1 atm displayed in Fig. 2 and those ones simulated in this study shows that all the coordination shells and subshells were well reproduced using the approach. We can also see in the simulated average $G(R)$ functions for the values $Kmax$ = 4 and 8 Å$^{-1}$, the subshells seen in Fig. 2 are not seen in the simulated average $G(R)$ functions due to their resolution $\Delta R$ = 0.95 Å ($Kmax$ = 4 Å$^{-1}$) and $\Delta R$ = 0.475 Å ($Kmax$ = 8 Å$^{-1}$) are very poor. On the other hand, the simulated average $G(R)$ functions for the values $Kmax$ = 15, 30, and 60 Å$^{-1}$ agree very well with the experimental one shown in Fig. 2. This is due to their high resolutions $\Delta R$ = 0.253 Å ($Kmax$ = 15 Å$^{-1}$), $\Delta R$ = 0.127 Å ($Kmax$ = 30 Å$^{-1}$), and $\Delta R$ = 0.063 Å ($Kmax$ = 60 Å$^{-1}$). It interesting to note that the simulated average $G(R)$ function for the value $Kmax$ = 60 Å$^{-1}$ shows that the coordination shells are formed by several subshells, which are not visible in the experimental $G(R)$ function.

Figures 3-6 show the simulated average $G_{Ta-Ta}(R)$, $G_{Ta-O}(R)$, and $G_{O-O}(R$ functions for the values $Kmax$ = 4, 8, 15, 30, and 60 Å$^{-1}$, respectively. Here, it is interesting to note that, usually, the PDF analysis for the crystalline materials do not get the $G_{ij}(R)$ function, and consequently, this fact is a significant advantage of using this approach. From these figures one can also see that the coordination shells of the $G_{ij}(R)$ functions are formed by several subshells.

The average partial radial distribution functions $RDF_{Ta-Ta}(R)$, $RDF_{Ta-O}(R)$, and $RDF_{O-O}(R$ functions were simulated for the values $Kmax$ = 4, 8, 15, 30, and 60 Å$^{-1}$, following the expression (11). Using the simulated average $G_{ij}(R)$ functions for $Kmax$ = 30 Å$^{-1}$ shown in Figs. 3-6, the lower and upper limits for the first coordination shells were obtained. These limits were considered to isolate the first coordination shells in the $RDF_{Ta-Ta}(R)$, $RDF_{Ta-O}(R)$, and $RDF_{O-O}(R$ functions in order to obtain the number of first neighbors, and they are shown in Fig. 7. By integrating these shells, the following values for the first neighbors were obtained: $N_{Ta-Ta}$ = 7.6 at $R$ = 3.87 Å, $N_{Ta-O}$ = 11.8 at $R$ = 1.96 Å, $N_{O-Ta}$ = 4.7 at $R$ = 3.87 Å, and $N_{O-O}$ = 19.2 at $R$ = 2.82 Å.

As mentioned in Ref. [8], with increasing pressure a pressure-induced amorphization process was observed. This amorphization can be associated with a chemical disorder promoted by the increase of pressure. Thus, it is interesting to examine this possibility. The Cowley-Warren chemical short-range order (CSRO) parameter is used to study the statistical distribution of atoms in solids, and is given by [17]

$$\alpha_{ij}^{CW} = 1.0 - \frac{N_{ij}}{c_j[c_j(N_{ii}+N_{ij})+c_i(N_{jj}+N_{ji})]},  \quad (12)$$

where $N_{ii}$, $N_{ij}$ and $N_{jj}$ are the coordination numbers and $c_i$ and $c_j$ are the concentrations of atoms of the elements $i$ and $j$. The $\alpha^{CW}$ parameter is zero for a random distribution, negative if there is a preference for forming unlike pairs and positive if homopolar pairs (clusters or local order) are preferred. Although the $\alpha^{CW}$ parameter is usually applied to amorphous phases, it can also be used to determine the relative preference for forming different atomic pairs in a crystalline binary alloy. Using the values $N_{Ta-Ta} = 7.6$, $N_{Ta-O} = 11.8$, $N_{O-Ta} = 4.7$, and $N_{O-O} = 19.2$ in Eq. (12), a value of $\alpha^{CW}{}_{Ta-O} = 0.201$ was calculated. This value suggests a preference for forming homopolar pairs in the first coordination shell. Thus, with increasing pressure, the repulsive part of the crystalline field plays an important role in the structural stability of the orthorhombic $Ta_2O_5$ phase. Then, an amorphous phase can be formed as reported by those researchers in Ref. [8].

## V. CONCLUSIONS

In this paper we have presented an approach combining the Rietveld method and pairs distribution function analysis to help understand the effects of temperature and/or high-pressure on crystalline materials. It was applied to orthorhombic $Ta_2O_5$ compound, and the results were compared with an experimental pairs distribution function $G(R)$ reported in the literature, and an excellent agreement was reached. In addition, the approach has permitted to simulate the average partial pairs distribution functions $G_{Ta-Ta}(R)$, $G_{Ta-O}(R)$, and $G_{O-O}(R$ and the average partial radial distribution functions $RDF_{Ta-Ta}(R)$, $RDF_{Ta-O}(R)$, and $RDF_{O-}$ $_O(R)$. From the latter, the coordination numbers for the first neighbors were obtained. They were used in the expression for the Cowley-Warren chemical short-range order (CSRO) parameter, and a value $\alpha^{CW} = 0.201$ was calculated, which suggests a preference for forming homopolar pairs in the first coordination shell. The reported pressure-induced amorphization process in orthorhombic $Ta_2O_5$ can be associated with the preference for forming homopolar pairs in the first coordination shell.

ACKNOWLEDGMENTS

One of the authors (Z. V. Borges) was financially supported by a scholarship from CNPq.

FIGURES

```
Space Group No.:            25                          Space Group No.:            25
Short Hermann-Mauguin Symbol: P M M 2                   Short Hermann-Mauguin Symbol: P M M 2
Schoenflies Symbol:         C2v1                        Schoenflies Symbol:         C2v1
a    =  43.99690                                        a    =  43.99690
b    =   3.89400                                        b    =   3.89400
c    =   6.20900                                        c    =   6.20900
alph =  90.00000                                        alph =  90.00000
beta =  90.00000                                        beta =  90.00000
gamma=  90.00000                                        gamma=  90.00000
(Length units: angstrom  Angle units: degree)           (Length units: angstrom  Angle units: degree)
********** Shell Structure **********                   ********** Shell Structure **********
Center: ( 3.80179e-019, 3.80179e-019, 0.006209)         Center: ( 3.78658e-016,  1.947,  6.18416)
Number of Atoms: 5301                                   Number of Atoms: 5263
Atom Symbol    x           y           z       distance from center    Atom Symbol    x           y           z       distance from center
   1 Ta     0.00000     0.00000     0.00621     0.00000                    1 O      0.00000     1.94700     6.18416     0.00000
   2 O      0.00000     0.00000     1.93721     1.93100                    2 Ta     0.00000     0.00000     6.21521     1.94725
   3 O      1.75988     0.00000    -0.81338     1.94136                    3 Ta     0.00000     3.89400     6.21521     1.94725
   4 O     -1.75988    -0.00000    -0.81338     1.94136                    4 O      1.75988     3.89400     5.39562     2.74040
   5 O     -0.00000    -1.94700    -0.02484     1.94725                    5 O      1.75988     0.00000     5.39562     2.74040
   6 O     -0.00000     1.94700    -0.02484     1.94725                    6 O     -1.75988     3.89400     5.39562     2.74040
   7 O     -0.00000    -0.00000    -2.02413     2.03034                    7 O     -1.75988     0.00000     5.39562     2.74040
   8 Ta    -1.89186    -0.00000    -2.88346     3.45389                    8 O      0.00000     3.89400     8.14621     2.76413
   9 Ta     1.89187    -0.00000    -2.88346     3.45389                    9 O      0.00000     0.00000     8.14621     2.76413
  10 Ta     3.45816     0.00000    -0.07451     3.45910                   10 O      0.00000     3.89400     4.18487     2.79070
                                                                          11 O      0.00000     0.00000     4.18487     2.79070
```

Figure 1: Output files of the Crystal Office 98® software listing the shell structures around a Tantalum atom at origin (left) and around an Oxygen atom at origin (right). The shell structures were calculated up to $R = 25$ Å, but they are shown up to $R \approx 3.46$ Å and $\approx 2.79$ Å.

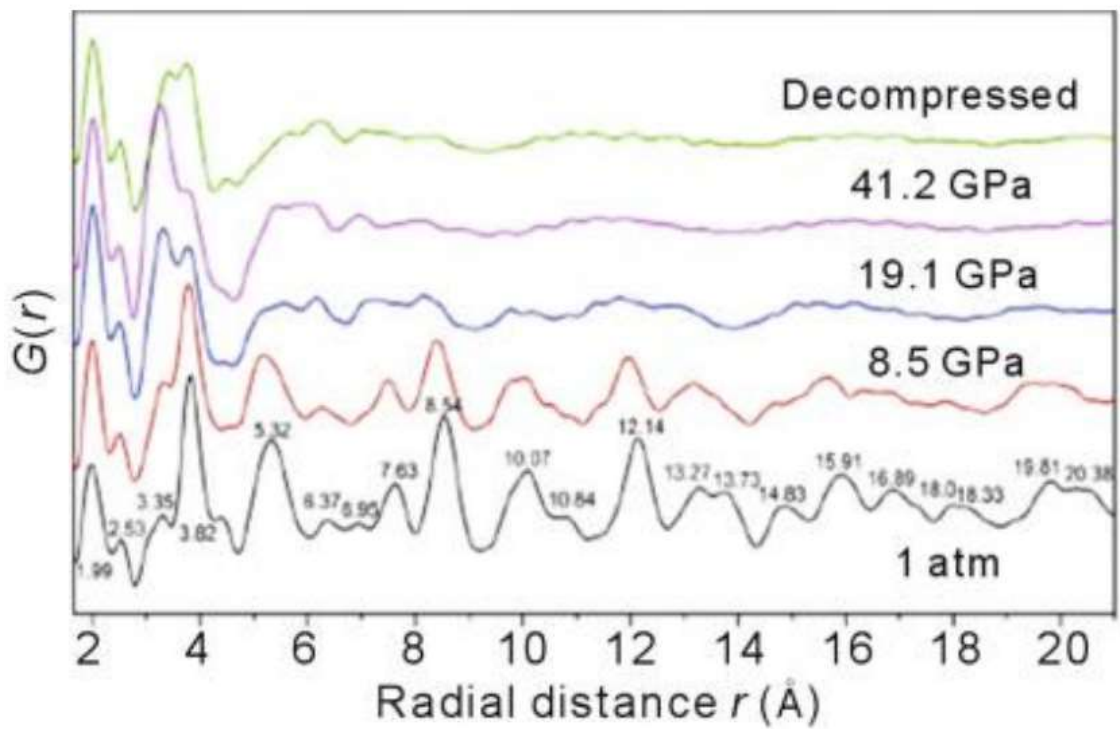

Figure 2: Experimental pairs distribution function $G(R)$ of orthorhombic $Ta_2O_5$ as a function of pressure applied. This figure was extracted from the X. Lu *et al.*, dx.doi.org/10.1021/ja407108u| J. Am. Chem. Soc.XXXX, XXX, XXX−XXX.

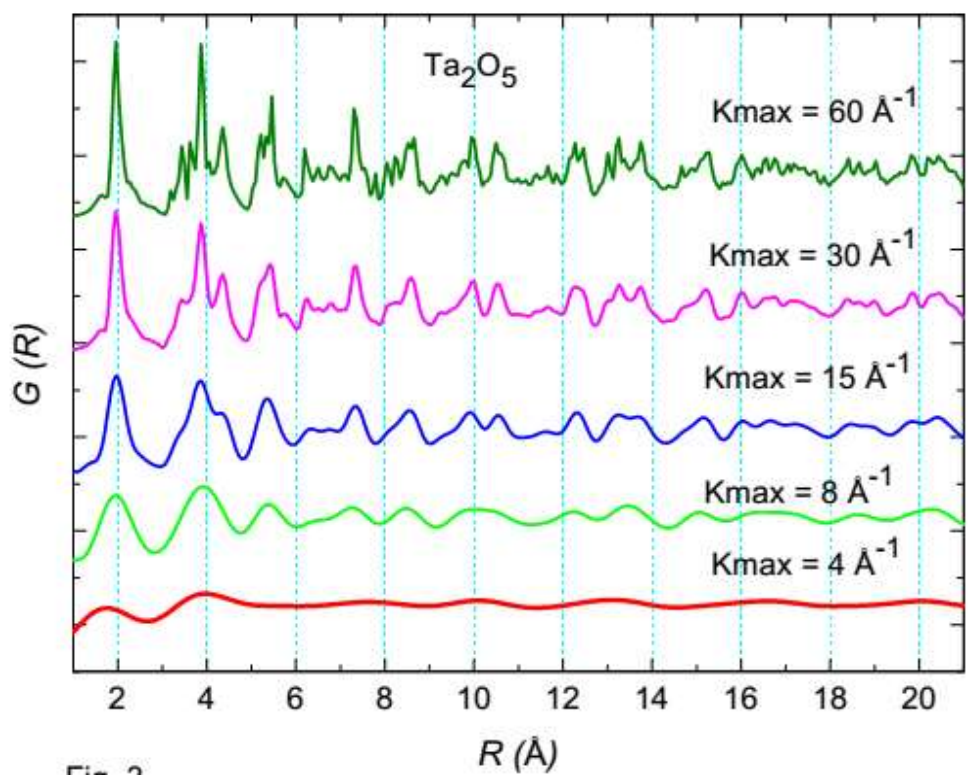

Figure 3: Simulated total pairs distribution function $G(R)$ for orthorhombic $Ta_2O_5$ with different values of the transferred momentum $K$ vector.

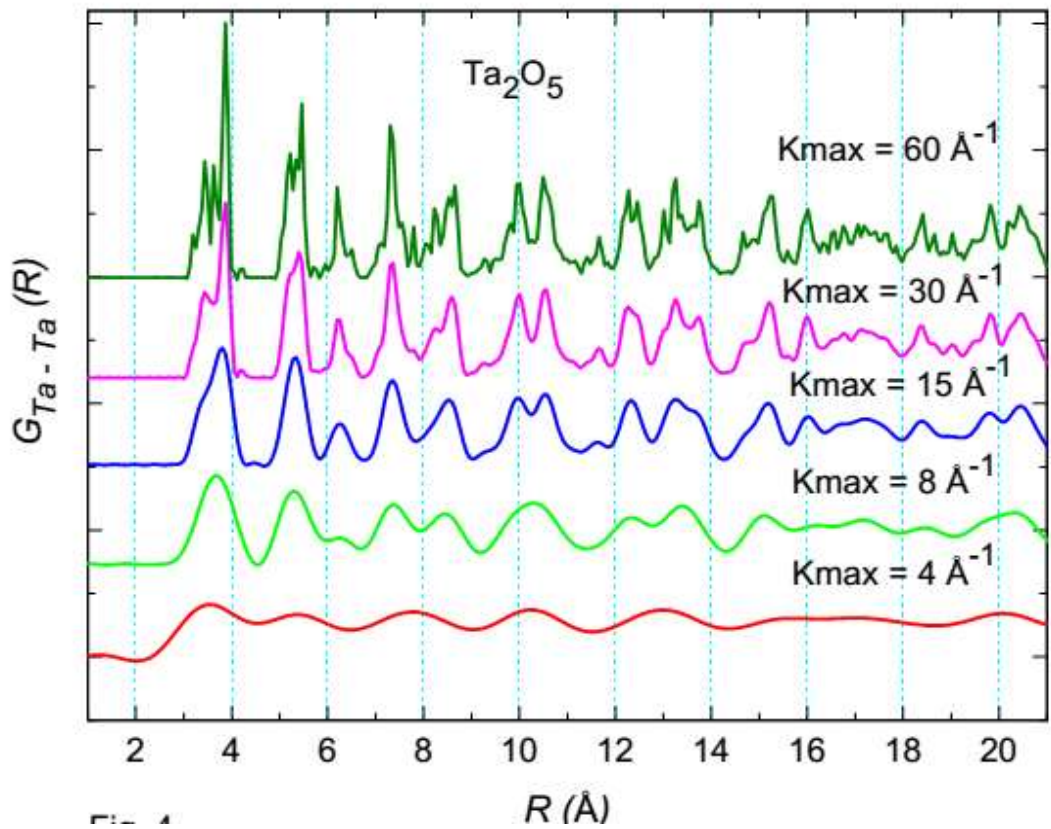

Figure 4: Simulated partial pairs distribution function $G_{Ta-Ta}(R)$ for orthorhombic $Ta_2O_5$ with different values of the transferred momentum $K$ vector.

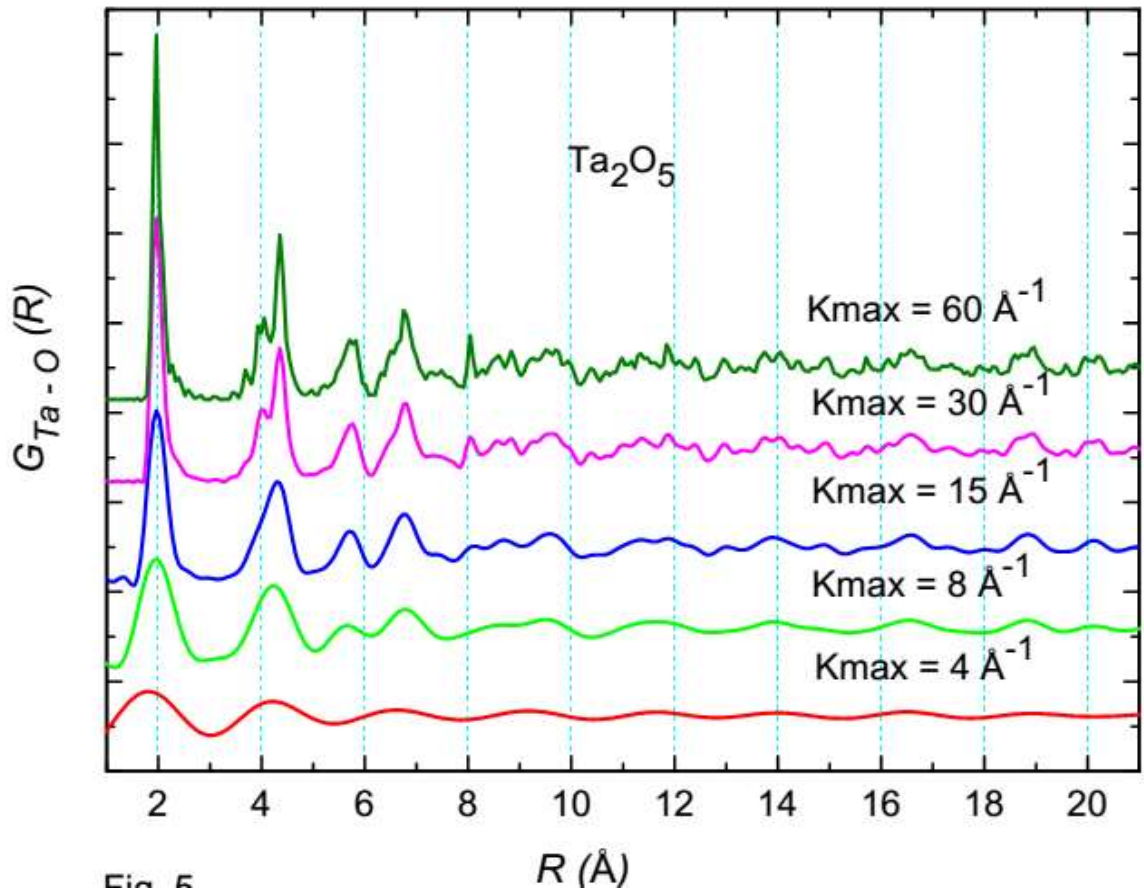

Figure 5: Simulated partial pairs distribution function $G_{Ta\text{-}O}(R)$ for orthorhombic $Ta_2O_5$ with different values of the transferred momentum $K$ vector.

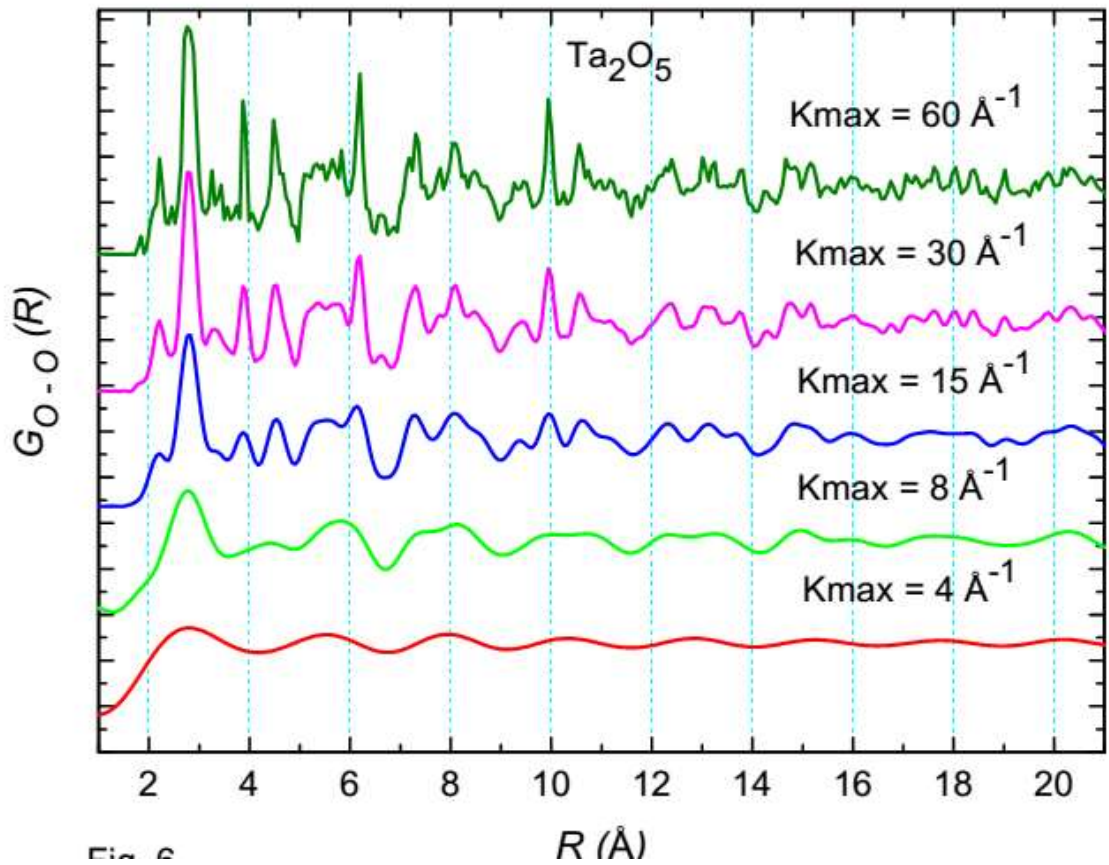

Figure 6: Simulated partial pairs distribution function $G_{O-O}(R)$ for orthorhombic $Ta_2O_5$ with different values of the transferred momentum $K$ vector.

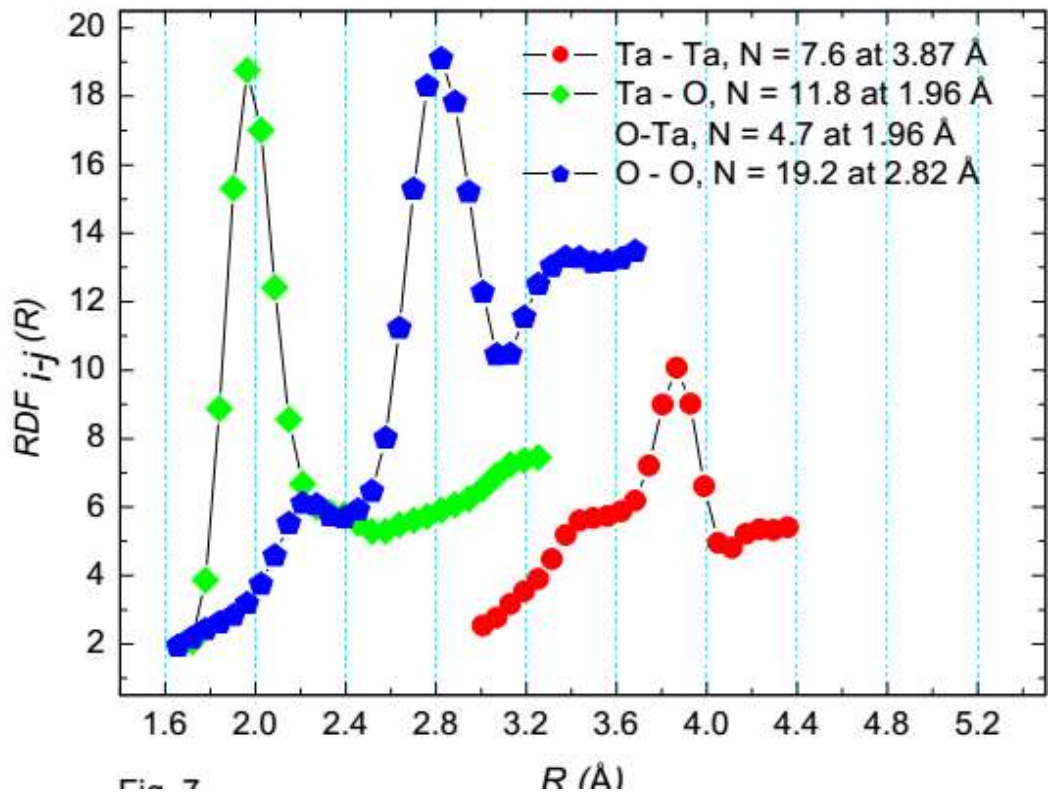

Figure 7: Simulated partial radial distribution functions $RDF_{i-j}(R)$ with the value $Kmax = 30$ Å$^{-1}$ for orthorhombic $Ta_2O_5$. Only the first coordination shells are shown.